# Quark Masses and Chiral Symmetry

Michael Creutz*

*Physics Department, Brookhaven National Laboratory, Upton, NY 11973; email: creutz@bnl.gov*

(May 10, 1995)

## Abstract

I discuss the global structure of the strongly interacting gauge theory of quarks and gluons as a function of the quark masses and the CP violating parameter $\theta$. I concentrate on whether a first order phase transition occurs at $\theta = \pi$. I show why this is expected when multiple flavors have a small degenerate mass. This transition can be removed by sufficient flavor-breaking. I speculate on the implications of this structure for Wilson's lattice fermions.

11.10.-z, 11.30.Rd, 11.15.Tk, 11.15.Ha

Typeset using REVTEX

# I. INTRODUCTION

This paper concerns the mass term $m\overline{\psi}\psi$ in the standard hadronic gauge theory of quarks and gluons. I was originally motivated by attempts to understand chiral symmetry on the lattice, but realized that my general understanding of chiral symmetry was inadequate. Thus arose the present discussion in a continuum framework. Indeed, this might be classified as a chiral-Lagrangian paper, in that I use primarily symmetry arguments to explore the theory as a function of the quark masses.

I will be quite cavalier about defining products of fermion fields, such as the mass term above. I assume that some regulator, such as the lattice, has made these quantities well defined. Some caution is needed with regard to issues such as the chiral anomalies, which arise because of the singular nature of products of fields at a single space-time point. I will, however, ignore other complications of the regulator, except when I explicitly conjecture on the role of the lattice doublers.

One of my goals is to provide an intuitive picture for the physical meaning of the CP-violating parameter of the strong interactions. This term, often called the $\theta$ term, is usually discussed in terms of topological excitations of the gauge fields. Here, however, I treat it entirely in terms of the chiral symmetries expected in the massless limit of the theory.

Among the conclusions is that a first-order transition is expected at $\theta = \pi$ when the flavors have a small but degenerate mass. This transition can be removed if flavor-breaking is large enough. At the transition, CP is spontaneously broken. I will also make a few remarks on the implications for the structure of Wilson's lattice fermions.

This is a subject with a long history, and much of what I say is buried in numerous previous studies. The implications of $\theta$ to the fermion mass matrix are well known to low-energy chiral Lagrangian discussions [1-8]. The occurrence of a first-order phase transition at large $\theta$ has been discussed in [2]. The possibility of a spontaneous breaking of CP was pointed out even before the significance of the parameter $\theta$ was appreciated [3]. The relation of $\theta$ to lattice Wilson fermions was elucidated some time ago by Seiler and Stamatescu [9] and was the subject of some recent work of my own [10]. My main new contributions are hopefully some added intuition to the understanding of these issues and more support for the structure of Wilson fermions presented in [10].

The sign of the fermion mass is sometimes regarded as a convention. Consider a Feynman diagram involving a fermion loop interacting with an arbitrary number of gauge boson lines. Insert a factor of $1 = (\gamma_5)^2$ at one vertex, and then move one of the factors of $\gamma_5$ around the loop, anti-commuting it with each gamma matrix it encounters. This reproduces the formal expression for the diagram but with each factor of $m$ replaced by $-m$. Thus we naively conclude that the physics of a gauge theory is unaltered by a change of the sign of the mass term.

This conclusion, however, is not true in general. It is probably correct for ordinary quantum electrodynamics in four space-time dimensions, where by Furry's theorem [11] there are no triangle diagrams and corresponding anomalies. However, it is explicitly false for the massive Schwinger model of electrodynamics in two space-time dimensions [12,10]. Furthermore, as the remaining discussion in this paper will argue, it is almost certainly true that hadronic physics would change if the sign of one of the quark masses were flipped.

To be a bit more general, consider a change of variables

$$\psi \longrightarrow e^{i\gamma_5 \theta/2}\psi. \tag{1}$$

Since $1 = (\gamma_5)^2$, this modifies the fermion mass term to

$$m\overline{\psi}\psi \longrightarrow m_1 \overline{\psi}\psi + im_2 \overline{\psi}\gamma_5\psi \tag{2}$$

where

$$\begin{aligned} m_1 &= m\cos(\theta) \\ m_2 &= m\sin(\theta). \end{aligned} \tag{3}$$

The kinetic and gauge terms of the gauge-theory action are formally invariant under this transformation. Thus, were one to start with the more general mass term of Eq. (2), one might expect a physical situation independent of $\theta$. However, because of the chiral anomaly, this is not true. The angle $\theta$ represents a further non-trivial parameter of the strong interactions, beyond the fermion masses. Its non-vanishing would give rise to CP violating processes. As such are not observed in hadronic physics, the numerical value of $\theta$ must be very small [5].

If Eq. (1) just represents a change of variables, how can this affect physics? The reason is entwined with the divergences of quantum field theory and the necessity of regularization. Fujikawa [13] has shown how the anomaly can be incorporated into the path integral formulation via the the fermionic measure, which becomes non-invariant under the above chiral rotation. More specifically, under a Pauli-Villars [14] approach $\theta$ represents a relative $\gamma_5$ rotation between the mass term for the fundamental particle and the mass term for the heavy regulator field. On the lattice with Wilson's fermion prescription [15], the doublers play this role of defining the relative chiral phase [9,10].

From this point of view, the fermion doublers of lattice gauge theory are not a nemesis, but rather are necessary to the physics of $\theta$ and the chiral anomaly. While the Wilson term does represent an explicit breaking of chiral symmetry, it is philosophically no worse than the heavy auxiliary field used in the Pauli-Villars approach.

Coleman [12] discussed the physics of this extra parameter in two dimensional electrodynamics, where it represents a background electric field. In Ref. [10] I used these results to infer a possible expected behavior of Wilson lattice fermions in both two and four dimensions. In particular I proposed generalized phase diagrams in the space of the parameters $m_1$ and $m_2$. In this paper I show how some of these features follow directly from chiral symmetries and details of the known particle spectrum. I frame the present discussion in the context of the continuum theory after any regulator has been removed.

The resulting diagrams are strongly dependent on the number of fermion flavors. With a single species, a first-order phase transition line runs down the negative $m_1$ axis, starting at a non zero value for $m_1$. This is sketched in Fig. (1). For two flavors the details depend on the sign of a term in the effective action, but I argue for two first-order phase transition lines, starting near the origin and running up and down the $m_2$ axis. For degenerate quarks these transitions meet at the chiral limit of vanishing fermion mass, while a small flavor breaking can separate the endpoints of these first-order lines. This is sketched in Fig. (2). The chiral limit is pinched between these endpoints. With $N_f > 2$ flavors, the argument is sharper, with the $(m_1, m_2)$ plane having $N_f$ first order phase transition lines all pointing at the origin. The conventionally normalized parameter $\theta$ is $N_f$ times the angle to a point in this plane, and these transition lines are each equivalent to $\theta$ going through $\pi$.

Whenever the number of flavors is odd, there is a first-order transition running down the negative $m_1$ axis. Along this line there is a spontaneous breaking of CP, with a natural order parameter being $\langle i\overline{\psi}\gamma_5\psi\rangle$. This possibility of a spontaneous breakdown of CP was noted some time ago by Dashen [3] and has reappeared at various times in the lattice context [16,17].

I begin my detailed discussion with the two flavor case. Here several simplifications make the physics particularly transparent. I then discuss how the one flavor result arises when one of these flavors is taken to a large mass. From this I conjecture an analogy with heavy doublers and Wilson lattice fermions. Finally, I discuss the general $N_f$ situation.

## II. TWO FLAVORS

I begin by defining eight fields around which the discussion revolves

$$
\begin{aligned}
\sigma &= c\overline{\psi}\psi \\
\vec{\pi} &= ic\overline{\psi}\gamma_5\vec{\tau}\psi \\
\eta &= ic\overline{\psi}\gamma_5\psi \\
\vec{\delta} &= c\overline{\psi}\vec{\tau}\psi.
\end{aligned}
\qquad (4)
$$

The fermion $\psi$ has two isospin components, for which $\vec{\tau}$ represents the standard Pauli matrices. The factor $c$ is inserted to give the fields their usual dimensions. Its value is not particularly relevant to the qualitative discussion that follows, but one convention is take $c = F/|\langle\overline{\psi}\psi\rangle|$ where $F$ is the pion decay constant and the condensate is in the standard vacuum.

Corresponding to each of these quantities is a physical spectrum. In some cases this is dominated by a known particle. There is the familiar triplet of pions around 140 MeV and the eta at 547 MeV. The others are not quite so clean, with a candidate for the isoscalar $\sigma$ being the $f_0(980)$ and for the isovector $\delta$ being the $a_0(980)$. These detailed identifications are not particularly important to the following discussion. One fact I emphasize is that the lightest particle in the $\delta$ channel appears to be heavier than the $\eta$.

Now consider an effective potential $V(\sigma,\vec{\pi},\eta,\vec{\delta})$ constructed for these fields. There are various formal ways of doing this, either from a Legendre transform on the generating function, or more physically by asking for the state of minimum energy given a set of expectation values for the fields. Two comments are in order. First, the fact that we are working with composite fields is inessential to the construction of $V$. Indeed, many of us still have qualms about distinguishing between elementary and composite fields. Second, formally the effective potential must be convex. Physically, this is because of a phase separation which would occur if an expectation value is held in an unstable region. Multiple minima are filled in by a Maxwell construction. This is a rather technical but well understood point which I ignore for the sake of pedagogy.

I first consider the theory with vanishing quark masses. I remind the reader that this limit is remarkable in that it is totally free of any adjustable parameters. In the continuum limit, the strong coupling constant is absorbed via the phenomenon of dimensional transmutation [18], and all dimensionless quantities are determined. In the full theory with the quark masses turned back on, the only parameters are those masses and $\theta$.

For the massless theory many of the chiral symmetries become exact. Because of the anomaly, the transformation of Eq. (1), which mixes the $\sigma$ and $\eta$ fields, is not a good symmetry. However flavored axial rotations should be valid. For example, the rotation

$$\psi \longrightarrow e^{i\gamma_5 \tau_3 \phi/2}\psi. \tag{5}$$

mixes $\sigma$ with $\pi_3$

$$\begin{aligned} \sigma &\longrightarrow +\cos(\phi)\sigma + \sin(\phi)\pi_3 \\ \pi_3 &\longrightarrow -\sin(\phi)\sigma + \cos(\phi)\pi_3 \end{aligned} \tag{6}$$

This transformation also mixes $\eta$ with $\delta_3$

$$\begin{aligned} \eta &\longrightarrow +\cos(\phi)\eta + \sin(\phi)\delta_3 \\ \delta_3 &\longrightarrow -\sin(\phi)\eta + \cos(\phi)\delta_3 \end{aligned} \tag{7}$$

For the massless theory, the effective potential must be invariant under such rotations. In this two flavor case, the consequences can be compactly expressed by going to a vector notation. I define the four component objects $\Sigma = (\sigma, \vec{\pi})$ and $\Delta = (\eta, \vec{\delta})$. Chiral symmetry then implies that the effective potential is a function only of invariants constructed from these four vectors. A complete set of invariants is $\{\Sigma^2, \Delta^2, \Sigma \cdot \Delta\}$.

This separation into two independent sets of fields is special to the two flavor case, but makes the behavior of the theory particularly transparent. When I turn to more flavors, the arguments must be modified, but the structure of the higher symmetries actually make the conclusions a bit stronger.

I now use the experimental fact that chiral symmetry appears to be spontaneously broken. The minimum of the effective potential should not occur for all fields having vanishing expectation. We also know that parity and flavor appear to be good symmetries of the strong interactions, and thus the expectation value of the fields can be chosen in the $\sigma$ direction. Temporarily ignoring the fields $\Delta$, I expect the potential to have the canonical "sombrero" shape, as would be stereotyped with the form

$$V = \lambda(\Sigma^2 - v^2)^2 = \lambda(\sigma^2 + \vec{\pi}^2 - v^2)^2 \tag{8}$$

Here $v$ is the magnitude of the vacuum expectation value for $\sigma$, and $\lambda$ is a coupling strength related to the $\sigma$ mass. The normalization convention mentioned below Eq. (4) would have $v = F/2$. I sketch the generic structure of the potential in Fig. (3). This gives the standard picture wherein pions are Goldstone bosons associated with fields oscillating along the degenerate minima of the potential.

Now consider the influence of the fields $\Delta$ on this potential. Taking only small values for these fields, I expand the potential about vanishing $\Delta$

$$V = \lambda(\Sigma^2 - v^2)^2 + \alpha\Delta^2 - \beta(\Sigma \cdot \Delta)^2 + \ldots \tag{9}$$

Here the coefficients $\alpha$ and $\beta$ are functions of $\Sigma^2$. For most purposes the value of the latter is approximately $v^2$; so, we can think of these coefficients as constants. Note also that since $\Sigma \cdot \Delta$ is odd under parity, the expansion in this quantity starts off quadratically.

The terms proportional to $\alpha$ and $\beta$ generate masses for the $\eta$ and $\delta$ particles. Since $\Delta^2 = \eta^2 + \vec{\delta}^2$, the $\alpha$ term gives them equal masses. Substituting $\Sigma \sim (v, \vec{0})$ gives $(\Sigma \cdot \Delta)^2 \sim v^2\eta^2$; thus, the $\beta$ term breaks the $\eta$–$\vec{\delta}$ degeneracy. Here is where the observation that the $\eta$ is lighter than the $\delta$ comes into play; I have written a minus sign in Eq. (9), thus making the expected sign of $\beta$ positive. I assume this for now, but will comment later on the changes that would occur were $\beta$ negative.

This discussion of the masses is not completely rigorous. I have ignored possible differences in the wave function renormalizations for the respective particles, and higher terms in the expansion of Eq. (9) could also affect the physical masses. However, all I really need for my later conclusions is the sign of the effective coupling between the $\Sigma$ and the $\Delta$ fields.

In conventional discussions the $\eta$ ($\eta'$ for 3 flavors) acquires its mass via topological excitations in the gauge fields. These effects are buried in the coefficients $\alpha$ and $\beta$. I only assume that these are generated appropriately, and make no comments on the detailed mechanisms.

Now I turn on the fermion masses. I consider small masses, and assume they appear as a general linear perturbation of the effective potential

$$V \longrightarrow V - (M_1 \cdot \Sigma + M_2 \cdot \Delta)/c. \tag{10}$$

Here the four-component objects $M_{1,2}$ represent the possible mass terms. The normalization constant $c$ appears in Eq. (4). The zeroth component of $M_1$ gives a conventional mass term proportional to $\overline{\psi}\psi$, contributing equally to both flavors. The mass splitting of the up and down quarks appears naturally in the third component of $M_2$, multiplying $\overline{\psi}\tau_3\psi$. The term $m_2$ of Eq. (2) lies in the zeroth component of $M_2$.

The two four vectors $M_{1,2}$ represent a total of 8 possible "mass terms." However, the chiral symmetries of the problem tell us that physics can only depend on invariants. For these I can take $M_1^2$, $M_2^2$, and $M_1 \cdot M_2$. That there are three parameters is reassuring; there are two quark masses $m_u$ and $m_d$ as well as the CP violating parameter $\theta$. The mapping between these parameterizations is non-linear, the conventional definitions giving

$$\begin{aligned} M_1^2 &= (m_u^2 + m_d^2)/4 + m_u m_d \cos(\theta)/2 \\ M_2^2 &= (m_u^2 + m_d^2)/4 - m_u m_d \cos(\theta)/2 \\ M_1 \cdot M_2 &= m_u m_d \sin(\theta)/2 \end{aligned} \tag{11}$$

Note if one of the quark masses, say $m_u$, vanishes, then the $\theta$ dependence drops out. While this may be a possible way to remove any unwanted CP violation from the strong interactions, having a single quark mass vanish represents a fine tuning which is not obviously more compelling than simply tuning $\theta$ to zero. Also, having $m_u = 0$ appears to be phenomenologically untenable [7,8].

I now give a physical picture of what the two mass terms $M_1$ and $M_2$ do to the "Mexican hat" structure of the massless potential. For $M_1$ this is easy; its simply tilts the sombrero. This is sketched in Fig. (4). The symmetry breaking is no longer spontaneous, with the tilt selecting the direction for $\Sigma$ field to acquire its expectation value. This picture is well known, giving rise to standard relations such as the square of the pion mass being linearly proportional to the quark mass [19].

The effect of $M_2$ is more subtle and represents my main interest here. This quantity has no direct coupling to the $\Sigma$ field; so, I must look for a higher order effect. The $M_2$ term represents a force pulling on the $\Delta$ field. It should give the latter an expectation value proportional to the strength, $\langle\Delta\rangle \propto M_2$. Once $\Delta$ gains an expectation value, it then effects $\Sigma$ through the $\alpha$ and $\beta$ terms of the potential in Eq. (9). The $\alpha$ term is a function only of $\Sigma^2$, and, at least for small $M_2$, should not qualitatively change the structure of the symmetry breaking. On the other hand, the $\beta$ term will warp the shape of our sombrero. As this term is quadratic in $\Sigma \cdot \Delta$, this warping is quadratic in the strength of $M_2$ and quadratic in $\Sigma$. With $\beta$ positive, as suggested above, this favors an expectation value of $\Sigma$ lying along the vector $M_2$, but the sign of this expectation is undetermined. This effect is sketched in Fig. (5).

To summarize, the effect of $M_1$ is to tilt our Mexican hat, while the effect of $M_2$ is to install a quadratic warping. The three parameters of the theory are the amount of tilt, the amount of warping, and, finally, the relative angle between these effects. To better understand the interplay of these various phenomena, I now consider three specific situations in more detail.

### III. CASE A: $M_1 || M_2$

First consider $M_1$ and $M_2$ parallel in the four vector sense. This is the situation when we have the two mass terms of Eq. (2) and no explicit breaking of flavor symmetry. Specifically, I take $M_1 = (m_1, \vec{0})$ and $M_2 = (m_2, \vec{0})$. In this case the warping and the tilting discussed in the last section are along the same axis.

Suppose I consider $m_2$ at some non-vanishing fixed value, and study the behavior of the vacuum as $m_1$ is varied. The $m_2$ term has warped the sombrero, but if $m_1$ is large enough, the potential will have a unique minimum in the direction of this pull. As $m_1$ is reduced in magnitude, the tilt decreases, and eventually the warping generates a second local minimum in the opposite sigma direction. As $m_1$ passes through zero, this second minimum becomes the lower of the two, and a first-order phase transition must occur exactly at $m_1 = 0$. This situation is sketched in Fig. (6). From Eq. (11) we see that this situation represents $m_u = m_d$ and $\theta = \pi$.

As $m_2$ decreases, the warping decreases, reducing the barrier between the two minima. This makes the transition softer. A small further perturbation in, say, the $\pi_3$ direction, will tilt the sombrero a bit to the side. If the warping is small enough, the field can then roll around the preferred side of the hat, thus opening a gap separating the positive $m_2$ phase transition line from that at negative $m_2$. In this way sufficient flavor breaking can remove the first-order phase transition at $\theta = \pi$. If I start at $\theta = 0$ with a mass splitting between the up and down quarks, an isoscalar chiral rotation to generate non-zero $\theta$ will generate just such a term.

Is the physical quark mass splitting sufficient to remove the $\theta = \pi$ transition? Given that this splitting is of comparable magnitude to the masses themselves, this is difficult to decide. Witten has argued [2], based on the largeness of the gauge group $SU(3)$, that the physical up-down mass difference is indeed too large for this transition to occur. On the other hand, it is a rather academic question, since changing $\theta$ is on the same footing as changing the quark masses, all of which have been predetermined by nature.

## IV. CASE B: $M_1 \perp M_2$

I now turn to a situation where $M_1$ and $M_2$ are orthogonal. To be specific, take $M_1 = (m_1, \vec{0})$ and $M_2 = (0, 0, 0, \delta m)$, which physically represents a flavor symmetric mass term $m_1 = (m_u + m_d)/2$ combined with a flavor breaking $\delta m = (m_d - m_u)/2$. Now $M_2$ warps the sombrero downwards in the $\pm \pi_3$ direction. A large $m_1$ would overcome this warping, still giving a vacuum with only $\sigma$ having an expectation value. However, as $m_1$ decreases in magnitude with a fixed $\delta m$, there eventually comes a point where the warping dominates the tilting. At this point we expect a new symmetry breaking to occur, with $\pi_3$ acquiring an expectation value. This is sketched in Fig. (7). As $\pi_3$ is a CP odd operator, this is a spontaneous breaking of CP. The possibility of such a spontaneous breaking of CP was pointed out some time ago by Dashen [3].

To make this into a proper two dimensional phase diagram, I add an $m_3 \pi_3$ piece to the potential. This effectively twists $M_1$ away from being exactly perpendicular to $M_2$. When such a term is present, it adds an explicit CP breaking term and can be expected to remove the transition, just as an applied field removes the phase transition in the Ising model. We thus have a phase diagram in the $(m_1, m_3)$ plane with a first-order transition connecting two symmetrically separated points on the $m_1$ axis. This is sketched in Fig. (8).

The endpoints of this transition line are associated with the two points where one of the respective quark masses vanishes. The phase transition occurs when the two flavors have masses of opposite sign. Simultaneously flipping the signs of both quark masses can always be done by a flavored chiral rotation, say about the $\pi_3$ axis, and thus is a good symmetry of the theory.

Taking one of the flavors to infinite mass provides a convenient way to understand the one flavor situation. As sketched in Fig. (8), this represents looking only at the vicinity of one endpoint of the above transition line. In terms of the light species, this transition represents a spontaneous breaking of CP with a non-vanishing expectation for $i\bar\psi \gamma_5 \psi$. In the lattice context the possibility of such a phase was mentioned briefly by Smit [16], and extensively discussed by Aoki and Gocksch [17].

## V. CASE C: $\beta < 0$

I now return to the case where $M_1$ is parallel to $M_2$, but consider the consequences were the parameter $\beta$ to be negative. As discussed above, this is unlikely to be the case for the continuum theory. Nonetheless, as I mention below, there is some evidence that this may be the case for the strongly coupled lattice model.

If $\beta$ is negative, then the warping will be upward rather than downward along the direction of $M_2$. The case of the $(m_1, m_2)$ plane, as discussed in section (III) for $M_1 \| M_2$, would at first glance seem to interchange with the picture of section (IV), for $M_1 \perp M_2$. Thus we expect an intermediate phase with a spontaneous generation of an expectation for the pion field. However, there is one crucial difference. In section (IV) I considered an explicit breaking of flavor, using the third direction for the example. In contrast, here I do not include any explicit flavor breaking. Thus when, say, $\pi_3$ acquires a vacuum expectation, it could just as well have been $\pi_1$ or $\pi_2$. For the pion to gain an expectation requires a spontaneous breaking of flavor symmetry. This is a continuous symmetry, and two Goldstone bosons appear.

The possibility of a phase with a spontaneous flavor breaking has been extensively discussed by Aoki and collaborators [17] in the context of Wilson lattice fermions at supercritical hopping. They give rather convincing arguments that such a phase indeed exists for strongly coupled lattice gauge fields. Similar phases have also been found for non-gauge models based on four fermion couplings [20].

Based on the prejudice that for two light flavors the physical value of the parameter $\beta$ should be negative, I suspect that these flavor violating phases are lattice artifacts. At strong coupling, $\beta$ may indeed be negative, but as the lattice spacing is reduced, it should switch to its physical sign and physics should revert to the situation of the previous sections. In the process the flavor breaking phase should be pinched out, much as seen in Ref. [20].

## VI. IMPLICATIONS FOR WILSON'S LATTICE FERMIONS

The Lagrangian for free Wilson lattice fermions is [15]

$$L(K, r, M) = \\ \sum_{j,\mu} K \left( \overline{\psi}_j (i\gamma_\mu + r) \psi_{j+e_\mu} + \overline{\psi}_{j+e_\mu} (-i\gamma_\mu + r) \psi_j \right) \\ + \sum_j (m_1 \overline{\psi}_j \psi_j + i m_2 \overline{\psi}_j \gamma_5 \psi_j) \qquad (12)$$

Here $j$ labels the sites of a four dimensional hyper-cubic lattice, $\mu$ runs over the space time directions, and $e_\mu$ is the unit vector in the $\mu$'th direction. I have scaled out all factors of the lattice spacing. The parameter $K$ is called the hopping parameter, and $r$ controls the strength of the so called "Wilson term," which separates off the famous doublers. I have also added to the theory of Ref. [15] an unconventional $m_2$ type mass term to make the connection with my earlier discussion.

Being quadratic and only involving nearest neighbor couplings, the spectrum is easily found by Fourier transformation. I omit the straightforward and well known details. Let me only observe that as a function of the mass parameter $m_1$, there are several places where the fermion spectrum has no mass gap. At these points there are massless particles in the spectrum. Conventionally, a massless fermion is obtained by taking $m_1 = 8Kr$, but there are other places where this original particle is massive while other doublers from the naive $r = 0$ theory become massless. At $m_1 = -8Kr$ one such species does so, at each of $m_1 = \pm 4Kr$ there are four massless doublers, and at $M = 0$ I find the remaining 6 of the total 16 species present in the naive theory.

My conjecture is that these various species should be thought of as flavors. When the gauge fields are turned on, the the full chiral structure should be a natural generalization of the earlier discussion. Thus near $m_1 = 8Kr$ I expect a first-order transition to end, much as is indicated in Fig. (1). This may join with numerous other transitions at the intermediate values of $m_1$, all of which then finally merge to give a single first-order transition line ending near $m_1 = -8Kr$. The situation near 0 and $\pm 4Kr$ involves larger numbers of flavors, and properly requires the analysis of the next section. However, one possible way the lines could join up is shown in Fig. (9a).

For two flavors of Wilson fermions, if we look near to the singularity at $8Kr$ we should obtain a picture similar to Fig. (2). However, further away these lines can curve and eventually end in the structure at the other doubling points. One possible picture is sketched in Fig. (9b). There may still be an Aoki phase, as discussed in the last section, appearing at strong coupling. But in the weak coupling limit I expect that phase to be squeezed out.

## VII. GENERAL $N_f$

As mentioned in the introduction, the general flavor case is somewhat more complicated because non-anomalous axial currents can mix the $\sigma$ and $\eta$ fields, making their separation less helpful. Given $N_f$ flavors, I turn to a matrix notation. My basic spin-zero fields are now

$$U_{ab} = c\overline{\psi}_a(1+\gamma_5)\psi_b. \tag{13}$$

The indices $a$ and $b$ run over the $N_f$ flavors, and $c$ is a normalization factor as introduced in Eq. (4).

Taking various traces will reproduce analogs of the fields introduced earlier. The $\sigma$ field is

$$\sigma = c\sum_a \overline{\psi}_a \psi_a = \mathrm{Tr}(U + U^\dagger). \tag{14}$$

The pseudo-scalar meson fields are

$$\pi_a = i\mathrm{Tr}(\lambda_a(U - U^\dagger)). \tag{15}$$

Here $\lambda_a$ represents a set of Hermitian and traceless generating matrices for $SU(N_f)$, i.e. extensions of the Pauli matrices for SU(2). For normalization, I choose

$$\mathrm{Tr}(\lambda_a \lambda_b) = 2\delta_{ab}. \tag{16}$$

The other relevant fields are generalizations of the $\eta$, now called $\eta'$, and $\delta$

$$\eta' = i\mathrm{Tr}(U - U^\dagger) \tag{17}$$

$$\delta_a = \mathrm{Tr}(\lambda_a(U + U^\dagger)). \tag{18}$$

The effective potential $V(U)$ is now a function of these $N_f^2$ fields. The consequences of chiral symmetry for the massless theory are that independent "left" and "right" $SU(N_f)$ rotations on the field $U$ do not change the potential. Thus I have

$$V(U) = V(g_L U g_R^{-1}) \tag{19}$$

whenever $g_L$ and $g_R$ are matrices from $SU(N_f)$. The theory has an $SU(N_f) \times SU(N_f)$ symmetry. Because of anomalies, $V(U)$ is not expected to be invariant under general phase changes. Thus Eq. (19) would not be true for arbitrary elements of $U(N_f) \times U(N_f)$. In particular, $V(U)$ can depend on the determinant of $U$.

I now assume that the massless theory undergoes chiral symmetry breaking in the usual manner; so, the field $U$ acquires a vacuum expectation value. Furthermore, I assume that flavor and parity are still good symmetries; so, I am free to pick this expectation value to be proportional to the unit matrix. This is analogous to taking the sigma direction for the breaking in the two flavor case. In this "standard" vacuum I have

$$\langle U_{ab} \rangle = v \delta_{ab} \tag{20}$$

Please don't confuse the Kronecker symbol here with the scalar field $\delta$ of before.

Chiral symmetry indicates that this choice is not unique; indeed, I could equivalently have chosen

$$\langle U_{ab} \rangle_g = v g_{ab} \tag{21}$$

where $g$ is an arbitrary element of $SU(N_f)$ and labels the respective vacuum. The set of degenerate vacua thus directly maps onto the group $SU(N_f)$. For two flavors this is the the four dimensional sphere ($S_3$), projected onto an circle for the earlier figures. One difference for higher $N_f$ is that anomaly free rotations, such as by nontrivial elements of the group center, can mix the $\sigma$ and $\eta'$ field. This is why the separation into fields $\Sigma$ and $\Delta$ is not generally useful.

I now add masses to perturb this manifold of vacua. Following the earlier discussion, the added terms are linear in the various scalar and pseudo-scalar quark bilinears

$$V(U) \longrightarrow V(U) + \mathrm{Tr}(MU + U^\dagger M^\dagger)/c. \tag{22}$$

Here the mass matrix $M$ is an arbitrary complex $N_f \times N_f$ quantity. Chiral symmetry says that physics is unchanged under taking $M \longrightarrow g_L M g_R^{-1}$, where $g_L$ and $g_R$ are elements of $SU(N_f)$. In this way an arbitrary mass matrix can be put into some standard form. Perhaps the most natural is the product of a real positive diagonal matrix, where the elements are the quark masses, with an overall phase factor $e^{i\theta/N_f}$.

To study the multi-flavor generalization of the $(m_1, m_2)$ phase diagram, I take all quark masses equal and let

$$M_{ab} = \delta_{ab}(m_1 + im_2). \tag{23}$$

What does such a distortion do to the manifold of ground states? The space of possible vacua is multi-dimensional, but by restricting the mass to this simple form, the important coordinates are the real and the imaginary parts of the trace of $U$. To help visualize the relevant structure, I generated 10,000 pseudo-random $SU(3)$ matrices with the invariant Haar measure, and plot in Fig. (10) the imaginary versus the real part of their traces. I also plot the boundary of the allowed region, mapped out as $t$ increases by matrices of the form $\exp(it\lambda_8)$, where $\lambda_8$ is chosen to be proportional to the diagonal matrix with elements $\{1, 1, -2\}$.

This figure dramatically shows the special role played by the center elements; these are of the form

$$g_{ab} = \delta_{ab} \exp(2\pi i n/N_f) \tag{24}$$

where $n$ is an integer. Fig. (11) contains a similar plot for $SU(4)$. This complex structure of the stationary points of the trace of an $SU(N)$ matrix is probably connected with the phase transitions seen at finite coupling for lattice gauge theory with $N$ larger than four [21].

The role of the group center can be understood more formally. The energy of the physical vacuum should be stationary under small variations of the expectation value of $U$. Making a variation in the group direction says that the vacuum must satisfy

$$0 = \left\langle \mathrm{Tr}(MU\lambda_a - \lambda_a U^\dagger M^\dagger) \right\rangle. \qquad (25)$$

Note that for the case of Eq. (23), where $M$ is proportional to the unit matrix, this equation is satisfied whenever $\langle U \rangle$ is itself proportional to the unit matrix. This is because the lambda matrices are traceless.

I assume that the perturbation is sufficiently small that the low energy states are still cleanly mapped onto the group $SU(N_f)$. Those elements of $SU(N_f)$ which are proportional to the unit matrix form the center of the group. Thus whenever $U$ is a constant times one of these elements, we have a candidate for the vacuum in the presence of the above mass term. While there are other stationary states in the group manifold, they are saddle points, and the true vacuum satisfies

$$\langle U \rangle = v \exp(2\pi i n / N_f) \qquad (26)$$

The mass addition can re scale the quantity $v$ from its value in the symmetry limit.

The diagonal mass term of Eq. (23) may be thought of as tilting the energy of the matrices in the $(\mathrm{Re}\,U, \mathrm{Im}\,U)$ plane. Depending on the direction of this tilt, the lowest energy state always lies at one of the vertices of the distorted $N_f$ polygon illustrated in Figs. (10,11). As the phase of the mass matrix changes, the preferred vacuum jumps from vertex to vertex. As we encircle the origin, we expect $N_f$ first-order phase transitions as we jump through these $N_f$ possible vacua, eventually returning to where we started. Each of the transitions is physically equivalent. The parameter $\theta$ is $N_f$ times the angle to the corresponding point in the $(m_1, m_2)$ plane; so, these transitions each correspond to $\theta$ going through $\pi$.

Note how for $N_f > 2$ the group itself contributes to warping the manifold of possible vacuum states. Whereas for two flavors I needed the term proportional to $\beta$ in the potential of Eq. (9), this is no longer the case with more flavors. The first-order transition at $\theta = \pi$ with many degenerate flavors comes naturally from the vacuum structure.

Just as in the two flavor case, I expect flavor breaking to complicate the picture. A more general mass term can pull in other directions in the group manifold, and the group center need no longer play a special role. In particular, this can cause the transitions near the origin to separate. Depending on the detailed masses, the case $\theta = \pi$ need no longer necessarily have a phase transition.

## VIII. SUMMARY AND CONCLUSIONS

I have presented a physical picture of the parameter $\theta$ in the context of an effective potential for spin-zero bilinears of the quark fields. I have argued for a first-order transition at $\theta = \pi$ when all flavors are degenerate, and shown how flavor breaking can remove this transition.

The picture may need to be modified for very large masses if the global structure of the effective potential is sufficiently modified. For example, a large $M_2$ could conceivably give a large enough contribution to the $\alpha$ term in Eq. (9) to destroy the spontaneous symmetry breaking on which the discussion is predicated. Then the first-order lines in Fig. (2) could end before reaching infinity.

The parameter $\theta$ is well known to generate $CP$ violation [1-5]. For example, the $\alpha$ term of Eq. (9) can give $\eta \to 2\pi$ decay when $\eta$ has an expectation value, as caused by the presence of $M_2$. In the context of my discussion, the strength of this decay is undetermined, but Refs. [5] make estimates using the full SU(3) symmetry. They also estimate the strength of the resulting neutron electric dipole moment, which is experimentally much more highly constrained. The unnaturally small value of $\theta$ is still unexplained.

A number of years ago Tudron and I [22] conjectured on the interplay of the confinement mechanism with $\theta$, and speculated that confinement might make $\theta$ unobservable. Recently Schierholz [23] argued that keeping confinement in the continuum limit may drive the theory to $\theta = 0$. The connection with present discussion is unclear, but the symmetries seem to indicate no obvious problem with $\theta$ being observable. Furthermore, the fact that the $\eta$ is lighter than particle candidates in the $\delta$ channel suggests that there indeed must be the $\beta$ term of Eq. (9), and it is this term which is directly responsible for the physical dependence on $\theta$.

An interesting question raised in Ref. [3] is whether the spontaneous parity violation appearing in these models might be responsible for the parity violation seen in nature. Generalizing these mass terms to Higgs couplings between gauged and spectator fermions should make it possible to create a mirror fermion model wherein the different helicities have different masses. Such models have been proposed [24] to circumvent difficulties of putting chiral fermions on the lattice [25]. While these are not ruled out experimentally, the necessity of adding the mirror states seems a bit artificial. Thus it would be interesting to know if there are any problems in principle with taking the auxiliary fermion masses to infinity. This raises triviality issues that deserve further study.

Finally, my discussion has been independent of conventional gluonic perturbation theory, and circumvents any perturbative mechanism for the $\theta$ dependence. Indeed, this phenomenon is non-perturbative and usually discussed in terms of topological effects [1]. The consequences, however, are highly constrained by confinement and the symmetries of the theory, allowing much to be said without specifying the underlying details.

FIGURES

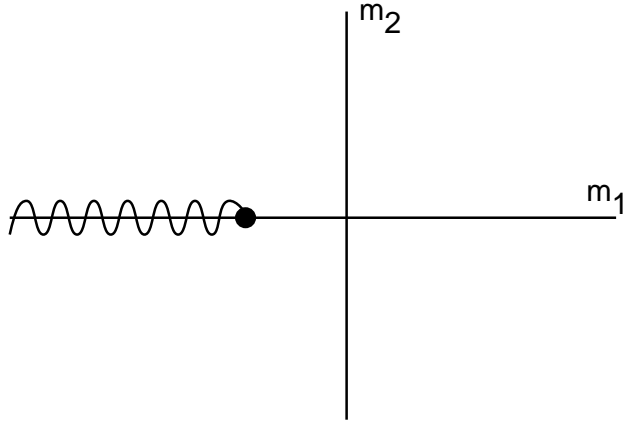

FIG. 1. The phase diagram for one flavor with the generalized mass term. The wavy line represents a first-order phase transition, along which $i\bar\psi\gamma_5\psi$ acquires an expectation value. The end point of this transition line is renormalized away from the origin towards negative $m_1$.

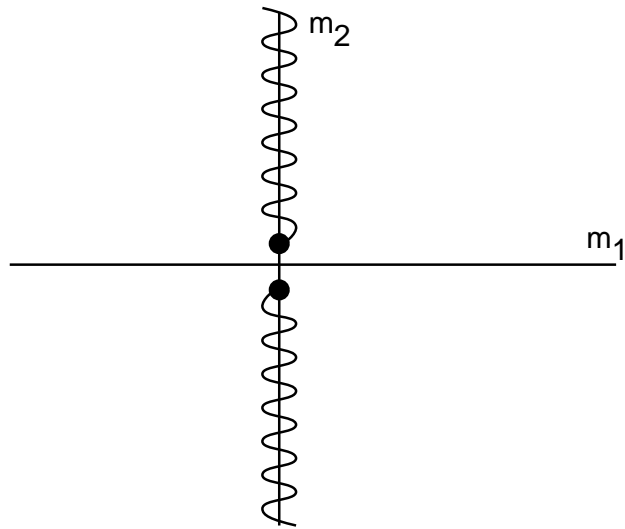

FIG. 2. The two flavor phase diagram. First-order lines run up and down the $m_2$ axis. The second order endpoints of these lines are separated by a flavor breaking mass difference. The chiral limit is pinched between these endpoints.

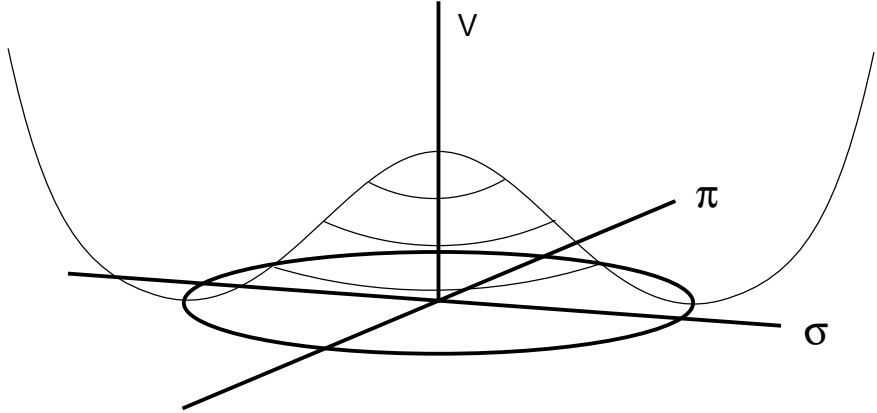

FIG. 3. The "sombrero" potential when the quark masses vanish.

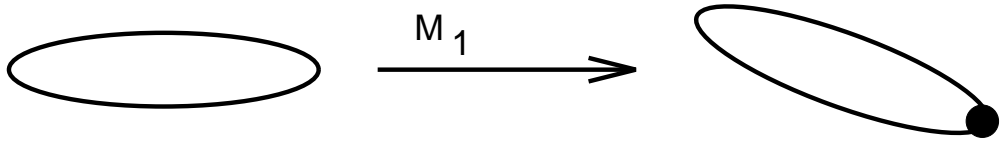

FIG. 4. The effect of $M_1$ on the effective potential. The ellipse in this and the following figures represents the minimum of the effective potential from Fig. (3). The dot represents where the vacuum settles.

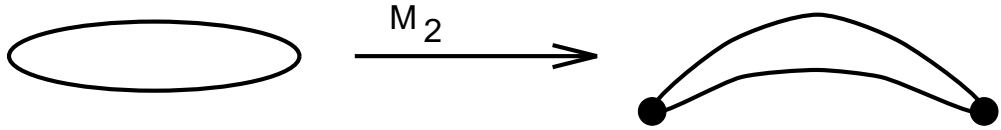

FIG. 5. The effect of $M_2$ on the effective potential. The dots represent two places where the vacuum can settle.

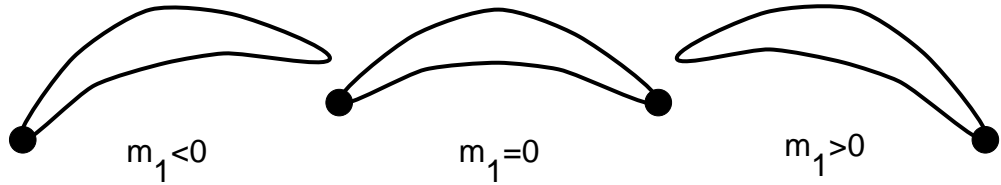

FIG. 6. Varying $m_1$ at fixed $m_2$. A first-order phase transition is expected at $m_1 = 0$. This corresponds to $\theta = \pi$. The dots represent places where the vacuum can settle.

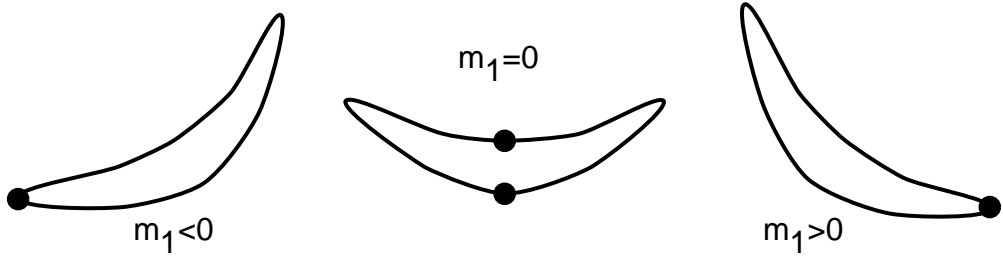

FIG. 7. Varying $m_1$ at fixed quark mass splitting. A second order phase transition occurs when the tilting is reduced sufficiently for a spontaneous expectation of $\pi_3$ to develop. The dots represent places where the vacuum can settle.

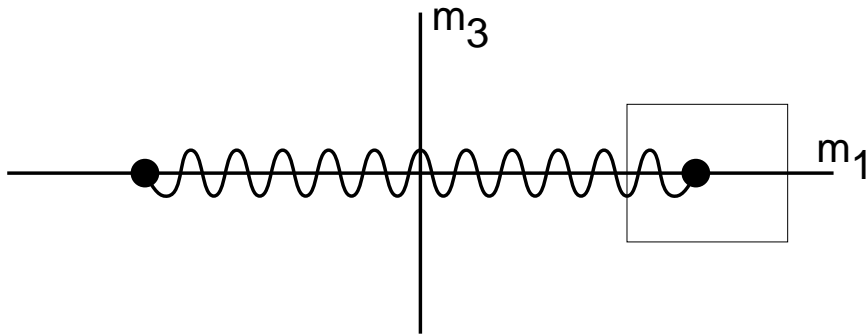

FIG. 8. The $(m_1, m_3)$ phase diagram for unequal mass quarks. The wavy line represents a first-order phase transition ending at the second order dots. The light box on the right shows how the one flavor diagram of Fig. (1) is extracted.

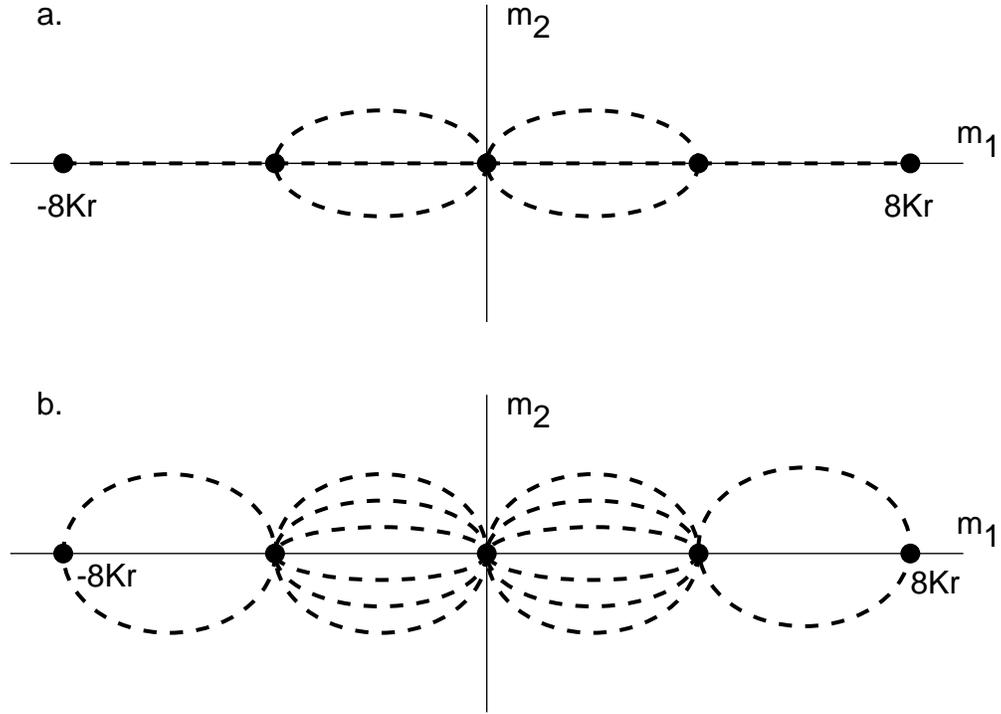

FIG. 9. Possible phase diagrams for lattice gauge theory with Wilson fermions. The dashed lines represent first-order phase transitions and the dots represent points where massless excitations should exist. Parts (a) and (b) are for the one and two flavor cases, respectively. The number of lines joining at each of these points counts the number of doubler species becoming massless there.

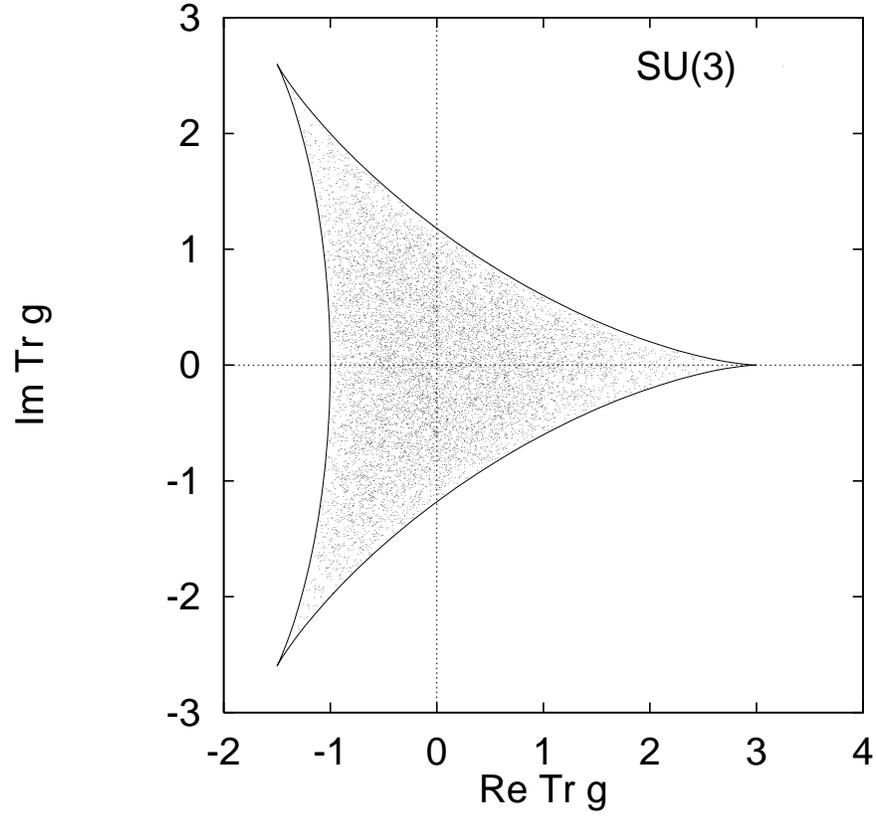

FIG. 10. Taking 10,000 $SU(3)$ matrices generated randomly with the invariant measure, I plot the real versus the imaginary parts of their traces. The boundary is obtained from matrices generated by $\lambda_8$.

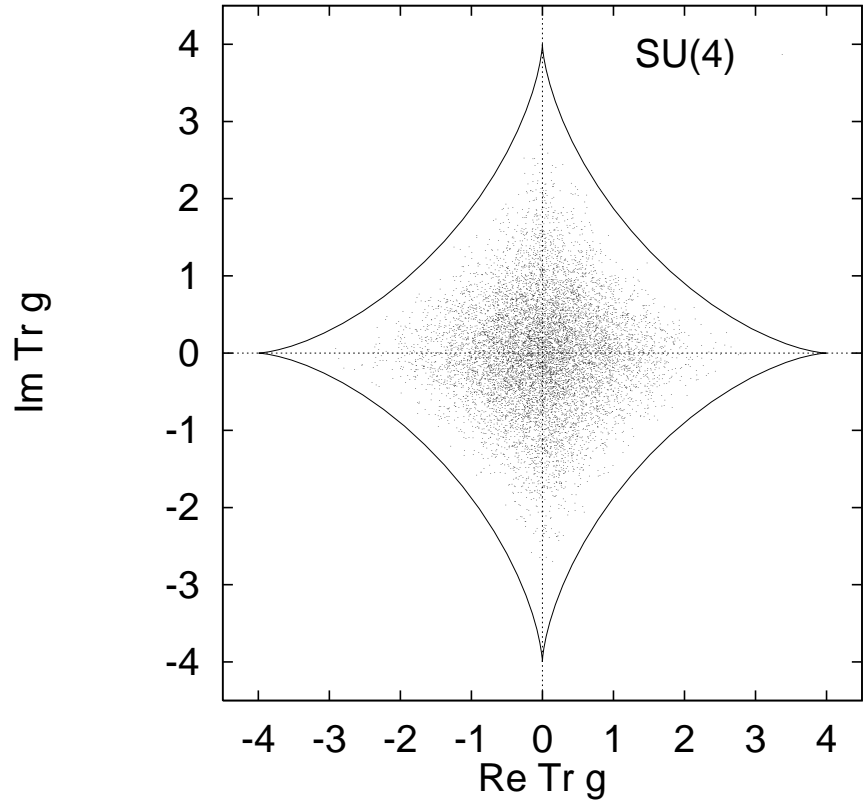

FIG. 11. Taking 10,000 $SU(4)$ matrices generated randomly with the invariant measure, I plot the real versus the imaginary parts of their traces. The boundary is obtained from diagonal matrices generated by $\text{diag}(1, 1, 1, -3)$.